# Universal Negative Correlation between Nonreciprocity and Superconducting Transition Width in FeTe-Based Heterostructures

Annie G. Wang[1,3], Saswata Mandal[1,3], Zi-Jie Yan[1,3], Bomin Zhang[1], Xiaoda Liu[1], Hongtao Rong[1], Deyi Zhuo[1], Pu Xiao[1], Lok-Kan Lai[1], Zhiyuan Xi[1], Ke Wang[2], Chao-Xing Liu[1], and Cui-Zu Chang[1]

[1]Department of Physics, The Pennsylvania State University, University Park, PA 16802, USA

[2]Materials Research Institute, The Pennsylvania State University, University Park, PA 16802, USA

[3]These authors contributed equally: Annie G. Wang, Saswata Mandal, and Zi-Jie Yan

Corresponding authors: cxl56@psu.edu (C.-X. L.); cxc955@psu.edu (C.-Z. C.).

**Abstract: Nonreciprocal charge transport in noncentrosymmetric superconductors probes broken symmetries and superconducting fluctuations and underpins the superconducting diode effect. Despite extensive studies, the factors governing its magnitude remain unclear. In this work, we use molecular beam epitaxy (MBE) to synthesize a broad family of FeTe-based superconducting heterostructures with top layers spanning topologically trivial versus nontrivial, layered versus nonlayered, and magnetic versus nonmagnetic materials. Through systematic first- and second-harmonic transport measurements, we find that the magnetochiral anisotropy coefficient $\gamma$ is insensitive to the top-layer crystal structure, lattice constant, thickness, topology, and magnetism, but exhibits a negative correlation with the superconducting transition width $\Delta T$. Control measurements on superconducting stoichiometric FeTe films reveal much weaker nonreciprocity, confirming the essential role of interfacial inversion-symmetry breaking. Combined with a disorder-based theoretical model, our results establish a universal relationship: for comparable inversion-symmetry**

**breaking, reduced interfacial disorder sharpens the superconducting transition and strongly enhances nonreciprocal transport in the Berezinskii-Kosterlitz-Thouless (BKT) transition regime. These findings identify $\Delta T$ as a unifying parameter governing nonreciprocity across FeTe-based heterostructures, reconcile previously disparate observations, and establish a general framework for engineering enhanced nonreciprocal superconducting responses and exploring superconducting diode effects in nonlinear quantum devices.**

**Main text:** Nonreciprocal charge transport is an asymmetric electrical response that depends on the current direction under an applied magnetic field and requires simultaneous breaking of inversion and time-reversal symmetries [1-6]. Early studies established this nonreciprocal transport phenomenon in nonsuperconducting materials, including noncentrosymmetric crystals [7-10], bulk chiral conductors [11], and topological materials [12-14]. More recently, nonreciprocal transport has emerged as an important probe of superconducting states, where Cooper-pair formation reduces the characteristic energy scale from the Fermi energy to the superconducting gap and enhances the magnetochiral anisotropy coefficient $\gamma$ by orders of magnitude [1,15-17]. Pronounced nonreciprocal charge transport has now been reported in a broad range of superconductors [15,18-34]. The strong enhancement of nonreciprocal response in the superconducting state has also stimulated intense interest in the superconducting diode effect, which has potential applications in low-power superconducting electronics and quantum computing [35-38].

FeTe-based heterostructures provide a versatile platform for investigating the origin of enhanced nonreciprocal charge transport. Interface-induced superconductivity has been established in FeTe-based heterostructures with Te-based topological insulators, including $Bi_2Te_3$ [18,39-41], $Sb_2Te_3$ [42], and $(Bi,Sb)_2Te_3$ [43,44]. A prior study attributed the thickness-dependent nonreciprocity in $Bi_2Te_3$/FeTe heterostructures to the interplay between topological surface states and superconductivity [18]. However, even stronger nonreciprocity was subsequently observed in topologically trivial ferromagnetic $CrTe_2$/FeTe heterostructures [19], demonstrating that topological surface states are *not* essential. These contrasting observations raise a fundamental question: what determines the magnitude of nonreciprocal charge transport in FeTe-based heterostructures? Although theoretical models have been proposed [15-17], the relationship between the magnitude of $\gamma$ and the quality of superconductivity in FeTe-based heterostructures with different top-layer materials remains poorly understood.

In this work, we establish a universal relationship between nonreciprocal charge transport and the superconducting transition width $\Delta T$ in FeTe-based heterostructures. Here, $\Delta T = T_{c,m} - T_{BKT}$ defines the temperature interval between the onset of mean-field superconductivity at $T_{c,m}$ [45] and the establishment of global phase coherence through the BKT transition at $T_{BKT}$ [46-48]. We systematically compare heterostructures with top layers spanning topologically trivial versus nontrivial, layered versus nonlayered, and magnetic versus nonmagnetic materials, including $Pb_{1-x}Sn_xTe$, $CrTe_2$, and $Bi_2Te_3$, as well as control superconducting stoichiometric FeTe films [49] and other superconducting FeTe-based heterostructures reported in prior studies [18,19,23]. Despite substantial differences in crystal structure, lattice constant, band topology, magnetism, and top-layer thickness, we find that none of these material properties primarily determines the magnitude of the nonreciprocal response. Instead, $\gamma$ exhibits a robust negative correlation with $\Delta T$: samples with a sharper superconducting transition, i.e., smaller $\Delta T$, consistently exhibit stronger nonreciprocal charge transport. This relationship persists across multiple SnTe/FeTe samples and the entire $Pb_{1-x}Sn_xTe$/FeTe series, and is further supported by $CrTe_2$/FeTe and $Bi_2Te_3$/FeTe heterostructures with different layer thicknesses. By incorporating quenched interfacial disorder into an inversion-breaking theoretical model describing nonlinear transport around the BKT transition, we show that disorder broadens $\Delta T$ and suppresses $\gamma$, thereby yielding a unified description of nonreciprocal charge transport in FeTe-based heterostructures.

All FeTe-based heterostructures and control superconducting FeTe films are grown on heat-treated $SrTiO_3$(100) substrates by MBE. Their structural properties are characterized by reflection high-energy electron diffraction (RHEED), X-ray diffraction (XRD), and scanning transmission electron microscopy (STEM). First- and second-harmonic longitudinal resistances $R^{\omega}$ and $R^{2\omega}$ are measured using lock-in amplifiers under an in-plane magnetic field perpendicular to the alternating current (AC). The magnetochiral anisotropy coefficient $\gamma$ is extracted from the antisymmetrized magnetic-field dependence of $R^{2\omega}/R^{\omega}$. The transition temperatures $T_{\mathrm{BKT}}$ and $T_{\mathrm{c,m}}$ are determined from Halperin-Nelson and Aslamazov-Larkin fits, respectively [45-48]. More details on the MBE growth, electrical transport measurements, and data analysis are provided in the Supplemental Material [50].

We first examine the relationship between $\gamma$ and $\Delta T$ in a series of 24-unit-cell (UC) $Pb_{1-x}Sn_xTe$/25-UC FeTe heterostructures with $0 \leq x \leq 1$, comprising rock-salt $Pb_{1-x}Sn_xTe$ on tetragonal FeTe (Fig. 1a). Their high crystalline quality is confirmed by sharp RHEED patterns (Fig. S1) [50] and XRD spectra [51]. We focus initially on three 24-UC SnTe/25-UC FeTe heterostructures, denoted Samples #1-#3, grown under different Sn/Te flux ratios. Despite comparable RHEED patterns (Fig. S2) [50], these samples exhibit distinct superconducting transitions (Figs. 1f and 2a,c). Cross-sectional STEM images of Sample #1 reveal well-ordered SnTe and FeTe lattices and an atomically sharp SnTe/FeTe interface without detectable elemental interdiffusion (Figs. 1b and S3) [50].

In superconducting FeTe-based heterostructures, inversion symmetry is inherently broken at the interface. Therefore, when an external in-plane magnetic field $\mu_0\boldsymbol{H}_{||}$ is applied to break time-reversal symmetry, a nonlinear contribution is expected to emerge in the longitudinal voltage $V_{xx}$ (Fig. 1c) [2,16,18,24].

$$V_{xx} = R^{\omega} I\,[1 + \gamma(\mu_0\boldsymbol{H}_{||} \times \hat{z}) \cdot \boldsymbol{I}], \quad (1)$$

where $\hat{z}$ denotes the surface-normal direction of the sample, and $\mu_0\boldsymbol{H}_{||}$ is applied parallel to the sample surface within the $x$-$y$ plane. The applied AC $\boldsymbol{I}$ flows along the $x$-axis, while $\varphi$ is defined as the angle between $\mu_0\boldsymbol{H}_{||}$ and the $y$-axis. Therefore, Equation (1) simplifies to $V_{xx} = R^{\omega}I + \gamma R^{\omega}\mu_0 H_{||} I^2 \cos\varphi$. With $I = \sqrt{2}\, I_0 \sin(\omega t)$, the second term gives rise to the second-harmonic longitudinal resistance $R^{2\omega} = \gamma R^{\omega}\mu_0 H_{||} I_0 \cos\varphi/\sqrt{2}$, from which $\gamma$ is extracted [50].

For Sample #1, Halperin-Nelson and Aslamazov-Larkin fits yield $T_{\mathrm{BKT}} \approx 11.62$ K and $T_{\mathrm{c,m}} \approx 13.00$ K, respectively, corresponding to $\Delta T = 1.38$ K (Figs. 1f and S4a) [45-48]. Within $T_{\mathrm{BKT}} < T < T_{\mathrm{c,m}}$, $R^{2\omega}$ exhibits a sinusoidal dependence on $\varphi$ whose amplitude increases with $\mu_0 H_{||}$ and reaches a maximum at $\varphi = 0°$ (Fig. 1d). Therefore, we measure $R^{\omega}$ and $R^{2\omega}$ at $\varphi = 0°$ and extract $\gamma$ from linear fits to $R^{2\omega}/R^{\omega}$ versus $\mu_0 H_{||}$, with a coefficient of determination of $R^2 \geq 0.99$ (Fig. 1e). As $T$ approaches $T_{\mathrm{BKT}}$, $\gamma$ increases sharply, reaching $\approx 23.00 \times 10^{-3}$ $\mathrm{A^{-1}T^{-1}m}$ at $T = 12.03$ K (Fig. 1g). This enhancement arises from the reduced superconducting energy scale and vortex-induced phase fluctuations [15,16,18,26]. For $T < T_{\mathrm{BKT}}$, both $R^{\omega}$ and $R^{2\omega}$ vanish, preventing the extraction of $\gamma$ from transport measurements [31].

Although the amplitude of the Cooper-pairing order parameter becomes nonzero for $T < T_{\mathrm{c,m}}$, phase fluctuations induced by thermally excited unbound vortices and antivortices continue to suppress the superfluid stiffness and sustain a finite resistance until the stiffness undergoes its universal jump at $T_{\mathrm{BKT}}$ [52]. Spatial inhomogeneity of the superconducting state can further modify electrical transport in this transition regime, thereby broadening $\Delta T$ [53,54]. To investigate the relationship between $\gamma$ and $\Delta T$, we study two additional 24-UC SnTe/25-UC FeTe heterostructures (Samples #2 and #3) with substantially larger $\Delta T$ than Sample #1.

Figure 2a,c shows the $R_{xx}$-$T$ curves and their corresponding Halperin-Nelson fits for Samples #2 and #3, respectively. Here, $R_{xx}$ denotes the sheet longitudinal resistance, defined as $R_{xx} = R^{\omega}/2$. For Sample #2, $T_{\mathrm{BKT}} \approx 8.70$ K and $T_{\mathrm{c,m}} \approx 12.40$ K, yielding $\Delta T = 3.70$ K. $\gamma$ reaches a maximum of $\approx 5.19 \times 10^{-3}$ $\mathrm{A^{-1}T^{-1}m}$ at $T = 9.50$ K (Fig. 2b). For Sample #3, $T_{\mathrm{BKT}} \approx 4.00$ K and $T_{\mathrm{c,m}} \approx 11.30$ K, yielding $\Delta T = 7.30$ K. $\gamma$ reaches a maximum of $\approx 2.37 \times 10^{-3}$ $\mathrm{A^{-1}T^{-1}m}$ at $T = 5.00$ K (Fig. 2d). Despite the substantially different $\Delta T$ values among Samples #1-#3, the three 24-UC SnTe/25-UC FeTe heterostructures exhibit a clear trend: the maximum $\gamma$ decreases as $T_{\mathrm{BKT}}$ decreases and $\Delta T$ broadens. We further investigate $m$-UC SnTe/40-UC FeTe heterostructures with $m = 3, 5, 9, 14, 19, 23$, and 24, all of which exhibit zero-resistance superconductivity [50]. Despite substantial variations in SnTe thickness, no systematic correlation is observed between $m$ and $\Delta T$. Instead, the maximum $\gamma$ generally decreases with increasing $\Delta T$ (Fig. S5 and Table S1) [50].

To establish the role of interfacial inversion-symmetry breaking, we investigate nonreciprocal charge transport in four stoichiometric FeTe films with varying thicknesses, which have recently been shown to exhibit superconductivity without an adjacent layer [49]. The 25-UC FeTe film exhibits $T_{\mathrm{BKT}} \approx 10.65$ K and $T_{\mathrm{c,m}} \approx 12.15$ K, yielding $\Delta T = 1.50$ K, comparable to that of Sample #1 (Fig. 2e). However, its maximum $\gamma$ is only $\approx 0.63 \times 10^{-3}$ $\mathrm{A^{-1}T^{-1}m}$ at $T = 11.20$ K (Fig. 2f), much lower than that of Sample #1 despite their comparable $\Delta T$. Similarly weak nonreciprocal responses are observed in the other three stoichiometric FeTe films with different thicknesses but comparable $\Delta T$ (Fig. S6 and Table S2) [50]. These results demonstrate that interfacial inversion-symmetry breaking is essential for generating the large values of $\gamma$ in FeTe-based heterostructures.

Prior studies have demonstrated that progressively substituting Pb for Sn in SnTe induces a topological phase transition from a topological crystalline insulator to a topologically trivial insulator [51,55-58]. To examine whether topology influences nonreciprocity, we investigate 24-UC $Pb_{1-x}Sn_xTe$/25-UC FeTe heterostructures with $x = 0$, 0.2, 0.4, 0.6, and 0.8 (Fig. 3a-e). Our recent angle-resolved photoemission spectroscopy measurements established a topological transition near $x = 0.72$, accompanied by a lattice-constant reduction from ~6.46 Å to ~6.35 Å [51]. Despite the topological phase transition and the accompanying decrease in lattice constant, all heterostructures exhibit $\Delta T < 2.5$ K, with maximum $\gamma$ values remaining within the same order of magnitude (Fig. 3 and Table S1) [50]. These findings indicate that neither top-layer topology nor lattice constant primarily determines the nonreciprocal response. The superconducting properties and variation in $\Delta T$ may instead be influenced by unintentional Te annealing of the underlying FeTe during top-layer growth [18,19,23,51].

Finally, we examine $CrTe_2$/FeTe and $Bi_2Te_3$/FeTe heterostructures, whose top layers are topologically trivial ferromagnetic and topological nonmagnetic materials, respectively. For a 4-trilayer (TL) $CrTe_2$/40-UC FeTe heterostructure, $\Delta T = 1.46$ K, and the maximum $\gamma$ reaches $\approx 39.62 \times 10^{-3}$ $A^{-1}T^{-1}m$ at $T = 11.50$ K [50]. Both $\Delta T$ and the maximum $\gamma$ are comparable to those of the 3-TL $CrTe_2$/40-UC FeTe and 5-TL $CrTe_2$/40-UC FeTe heterostructures (Fig. S7) [19,50]. To further test whether $\gamma$ is governed by $\Delta T$ rather than FeTe thickness, we measure another $CrTe_2$/FeTe heterostructure with a different FeTe thickness: 10-TL $CrTe_2$/60-UC FeTe (Fig. S8) [50]. This heterostructure exhibits $\Delta T$ = 1.97 K and a maximum $\gamma \approx 16.61 \times 10^{-3}$ $A^{-1}T^{-1}m$ at $T$ = 11.10 K. Both $\Delta T$ and the maximum $\gamma$ are also comparable to those of the 10-TL $CrTe_2$/20-UC FeTe heterostructure [19]. These results further indicate that $\gamma$ correlates with $\Delta T$ rather than with the thickness of either constituent layer. We further measure two $Bi_2Te_3$/FeTe heterostructures: 5-quintuple-layer (QL) $Bi_2Te_3$/40-UC FeTe and 8-QL $Bi_2Te_3$/50-UC FeTe heterostructures (Fig. S9) [50]. They exhibit $\Delta T$ = 3.06 K and 3.14 K, respectively, both smaller than those for $Bi_2Te_3$/FeTe heterostructures in a prior study [18]. Correspondingly, their $\gamma$ values reach $\approx 12.00 \times 10^{-3}$ $A^{-1}T^{-1}m$ and $17.50 \times 10^{-3}$ $A^{-1}T^{-1}m$ at $T$ = 9.30 K and 9.80 K, respectively, larger than those reported previously for $Bi_2Te_3$/FeTe heterostructures [18]. Therefore, despite differences in $Bi_2Te_3$ and FeTe layer thicknesses, heterostructures with smaller $\Delta T$ also consistently show larger $\gamma$, further supporting the negative correlation between nonreciprocity and $\Delta T$.

Figure 4a summarizes $\gamma$ values as a function of $\Delta T$ across all studied $Pb_{1-x}Sn_xTe$/FeTe, $CrTe_2$/FeTe, and $Bi_2Te_3$/FeTe heterostructures, four control superconducting FeTe films, and additional superconducting FeTe-based heterostructures reported in prior studies [18,19,23]. The data points for the control superconducting FeTe films deviate substantially from those of the FeTe-based heterostructures, which we attribute to the absence of a top layer and the associated interfacial inversion-symmetry breaking. Taken together, these results across $Pb_{1-x}Sn_xTe$/FeTe, $CrTe_2$/FeTe, $Bi_2Te_3$/FeTe, and $ZrTe_2$/FeTe heterostructures [18,19,23] demonstrate that $\Delta T$, rather than the specific top-layer material or layer thickness, is the dominant parameter governing the magnitude of $\gamma$ in FeTe-based heterostructures.

To understand the correlation between $\gamma$ and $\Delta T$ in superconducting FeTe-based heterostructures, we consider a Ginzburg-Landau (GL) free energy incorporating a magnetoelectric term and quenched disorder,

$$\mathcal{F}_{GL} = \int d^2\mathbf{r}\left[a(\mathbf{r})|\Psi(\mathbf{r})|^2 + \frac{b}{2}|\Psi(\mathbf{r})|^4 + \frac{1}{2m^*}|\nabla\Psi(\mathbf{r})|^2 - \frac{i\Lambda}{m^*}\{\Psi^*(\mathbf{r})\nabla^2(\boldsymbol{B}\times\nabla)\cdot\hat{z}\}\Psi(\mathbf{r})\right]. \quad (2)$$

Here, $\Psi(\boldsymbol{r})$ is the pairing function at the interface, $a(\boldsymbol{r})$ and $b$ are the GL coefficients, and $m^*$ is the Cooper-pair mass. The $\Lambda$ term describes the magnetoelectric coupling between the supercurrent and magnetic field $\boldsymbol{B} = \mu_0\boldsymbol{H}_{||}$, arising from broken inversion symmetry at the interface [16]. Quenched disorder produces superconducting regions with different critical temperatures [54,59-61], modeled by $a(\boldsymbol{r}) = \bar{a} + \delta a(\boldsymbol{r})$, where $\bar{a}$ gives the spatial average and $\delta a(\boldsymbol{r})$ represents Gaussian white-noise fluctuations satisfying $\langle\delta a(\boldsymbol{r})\rangle = 0$ and $\langle\delta a(\boldsymbol{r})\delta a(\boldsymbol{r}')\rangle = W\delta^{(2)}(\boldsymbol{r}-\boldsymbol{r}')$, with $W$ denoting the disorder strength. As derived in the Supplemental Material [50], averaging over the disorder $\delta a(\boldsymbol{r})$ with Gaussian distribution raises the mean-field critical temperature $T_{c,m}$ [53,54] and lowers the BKT temperature $T_{BKT}$ [52,62], thereby broadening $\Delta T$:

$$\Delta T \approx \Delta T_{clean} + \frac{m^* W}{\pi\alpha}\ln\frac{L}{a_0}, \quad (3)$$

where $\Delta T_{clean}$ is the superconducting transition width in the clean limit, $\alpha = \bar{a}/(T - T_{c,m})$, and $L$ and $a_0$ are the infrared and ultraviolet cutoff lengths, respectively, that regularize the logarithmic disorder correction in 2D. The magnetoelectric $\Lambda$ term couples the in-plane magnetic field $\mu_0\boldsymbol{H}_{||}$ to the supercurrent, introducing a current-direction-dependent correction to the superfluid stiffness. In the BKT regime, where resistance originates from the motion of thermally unbound vortices, this correction shifts $T_{BKT}$, and consequently the free-vortex density, in opposite directions for opposite currents. The strong $T$ dependence of the free-vortex density near $T_{BKT}$ enhances the nonreciprocal vortex-flow resistivity, characterized by the magnetochiral anisotropy coefficient $\gamma$. In contrast, disorder suppresses the superfluid stiffness, reducing the current-induced shift of $T_{BKT}$ and weakening nonreciprocity. As derived in the Supplemental Material [50], $\gamma$ takes the form

$$(T - T_{\mathrm{BKT}})^{1.5}\gamma(\Delta T) = \sqrt{\beta}\,\frac{2\pi}{e^*}\,\Lambda\left[\sqrt{\Delta T} - \frac{(\Delta T - \Delta T_{\mathrm{clean}})}{2\ln\left(\frac{L}{a_0}\right)\sqrt{\Delta T}}\left\{\ln(1 + 4m^*\alpha\Delta T L^2) + \frac{3}{2}\right\}\right]. \quad (4)$$

Here, $e^*$ is the Cooper-pair charge and $\beta$ is a nonuniversal constant of order unity from the Halperin-Nelson renormalization-group analysis of the BKT transition [50]. In the clean limit, i.e., $\Delta T \to \Delta T_{\mathrm{clean}}$, Equation (4) reduces to $\gamma \propto \frac{\sqrt{\Delta T_{\mathrm{clean}}}}{(T - T_{\mathrm{BKT}})^{1.5}}$, consistent with prior theoretical predictions [16]. Figure 4b compares the measured $(T - T_{\mathrm{BKT}})^{1.5}\gamma$ as a function of $\Delta T$ with the theoretical prediction (black curve) from Eq. (4), using $\frac{\Lambda\sqrt{\beta}}{e^*} = 1.25\ \mathrm{mA}^{-1}\mathrm{T}^{-1}$, $\alpha = 0.41\frac{J}{K}$, $L = \frac{11.2}{\sqrt{m^* k_B}}\hbar\ \mathrm{m} \approx 47a_0$, and $\Delta T_{\mathrm{clean}} = 1.38\ \mathrm{K}$. Here, $\Delta T_{\mathrm{clean}}$ is determined from the narrowest transition width observed among all our samples. The agreement across the FeTe-based heterostructures supports a universal disorder-controlled relationship between $\gamma$ and $\Delta T$, in which stronger interfacial disorder broadens $\Delta T$ and suppresses $\gamma$.

To summarize, we establish a universal negative correlation between the magnetochiral anisotropy coefficient $\gamma$ and the superconducting transition width $\Delta T$ across a broad family of FeTe-based heterostructures, independent of the top-layer crystal structure, lattice constant, topology, magnetism, or thickness. Our disorder-based BKT model reveals that quenched interfacial disorder broadens $\Delta T$ and suppresses $\gamma$ by depleting the superfluid stiffness. While interfacial inversion-symmetry breaking enables nonreciprocity, interfacial disorder limits its magnitude. These findings identify $\Delta T$ as a practical parameter for optimizing nonreciprocal superconducting transport and highlight interfacial quality, rather than the choice of top-layer material, as a key consideration in engineering superconducting diode effects.

**Acknowledgments:** We thank S. Islam, N. Samarth, B. Yan, and J. Zhu for helpful discussions. This project is primarily supported by the Office of Naval Research grant (N000142412133), including MBE growth and all PPMS measurements. Sample characterization is supported by the U.S. Department of Energy grant (DE-SC0023113) and the Gordon and Betty Moore Foundation (GBMF14458, DOI: 10.37807/GBMF14458). Theoretical calculations are partially supported by the National Science Foundation grant (DMR-2241327). C.-Z. C. acknowledges the support from the Gordon and Betty Moore Foundation's EPiQS Initiative (GBMF9063 to C.-Z. C.).

**Data availability**: The data that support the findings of this article are openly available [63].

**Figures and figure captions:**

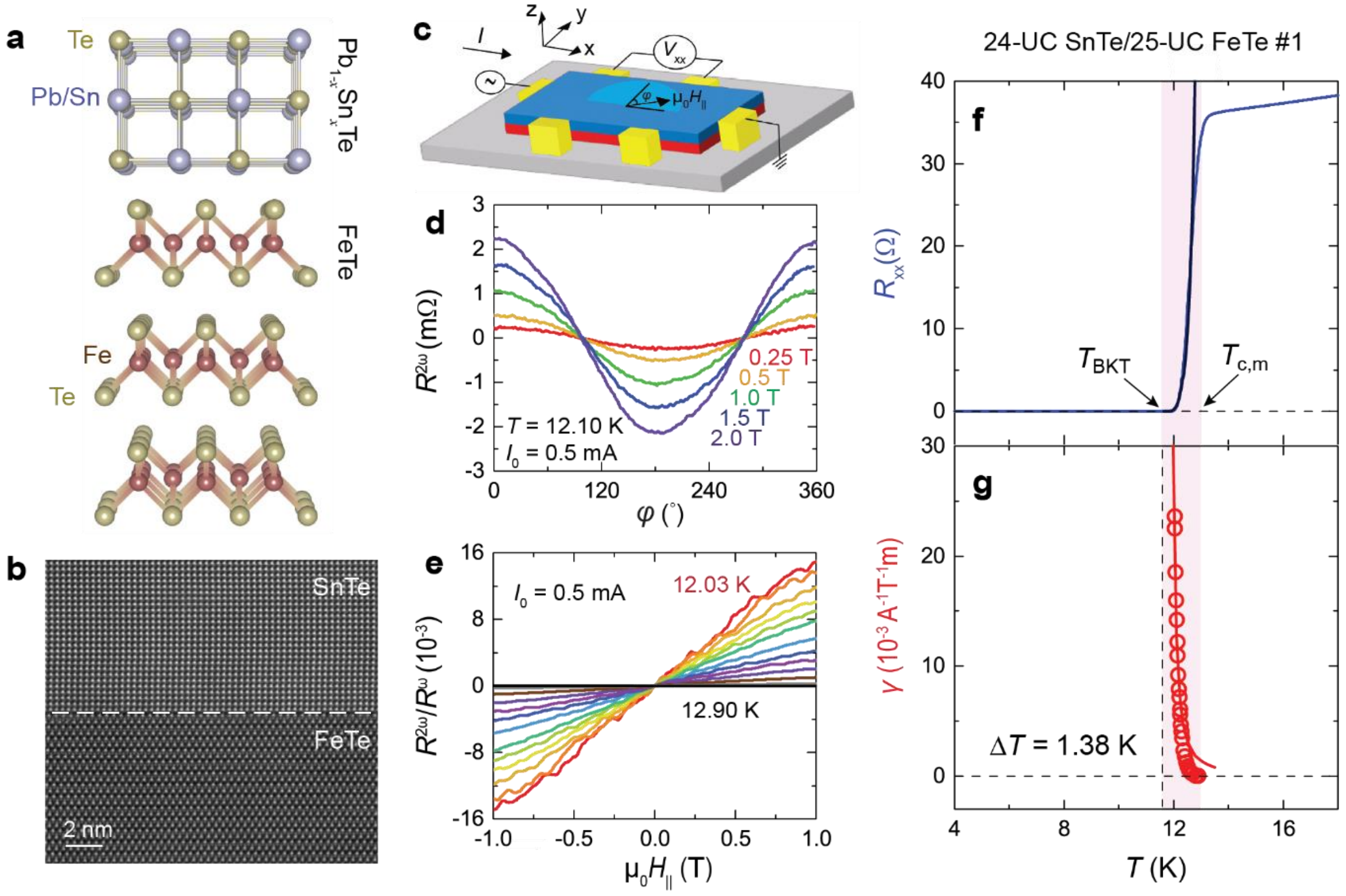

**Fig. 1 | Nonreciprocal charge transport in a 24-UC SnTe/25-UC FeTe heterostructure (Sample #1)**. **a,** Schematic lattice structure of the $Pb_{1-x}Sn_xTe$/FeTe heterostructure. **b,** Cross-sectional STEM image of a 24-UC SnTe/25-UC FeTe heterostructure. **c**, Schematic of the Hall-bar device used for nonreciprocal charge-transport measurements. $\varphi$ is defined as the angle between $\mu_0 H_{||}$ and the $y$-axis. **d**, $\varphi$-dependent $R^{2\omega}$ measured at $T = 12.10$ K and $I_0 = 0.5$ mA under different $\mu_0 H_{||}$. **e**, $\mu_0 H_{||}$-dependent $R^{2\omega}/R^{\omega}$ measured at $\varphi = 0°$ and 12.03 K $\leq T \leq$ 12.90 K. **f**, $T$-dependent sheet longitudinal resistance $R_{xx}$. The black curve is a fit using the Halperin-Nelson formula, yielding a BKT transition temperature $T_{BKT} \approx 11.62$ K. **g**, $T$-dependent $\gamma$. The red curve is a fit to $\gamma = \gamma_0 (T - T_{BKT})^{-1.5}$. The pink shaded regions in (**f**, **g**) indicate the temperature range of $T_{BKT} \leq T \leq T_{c,m}$, where $T_{c,m}$ is the mean-field transition temperature determined from the Aslamazov-Larkin fit.

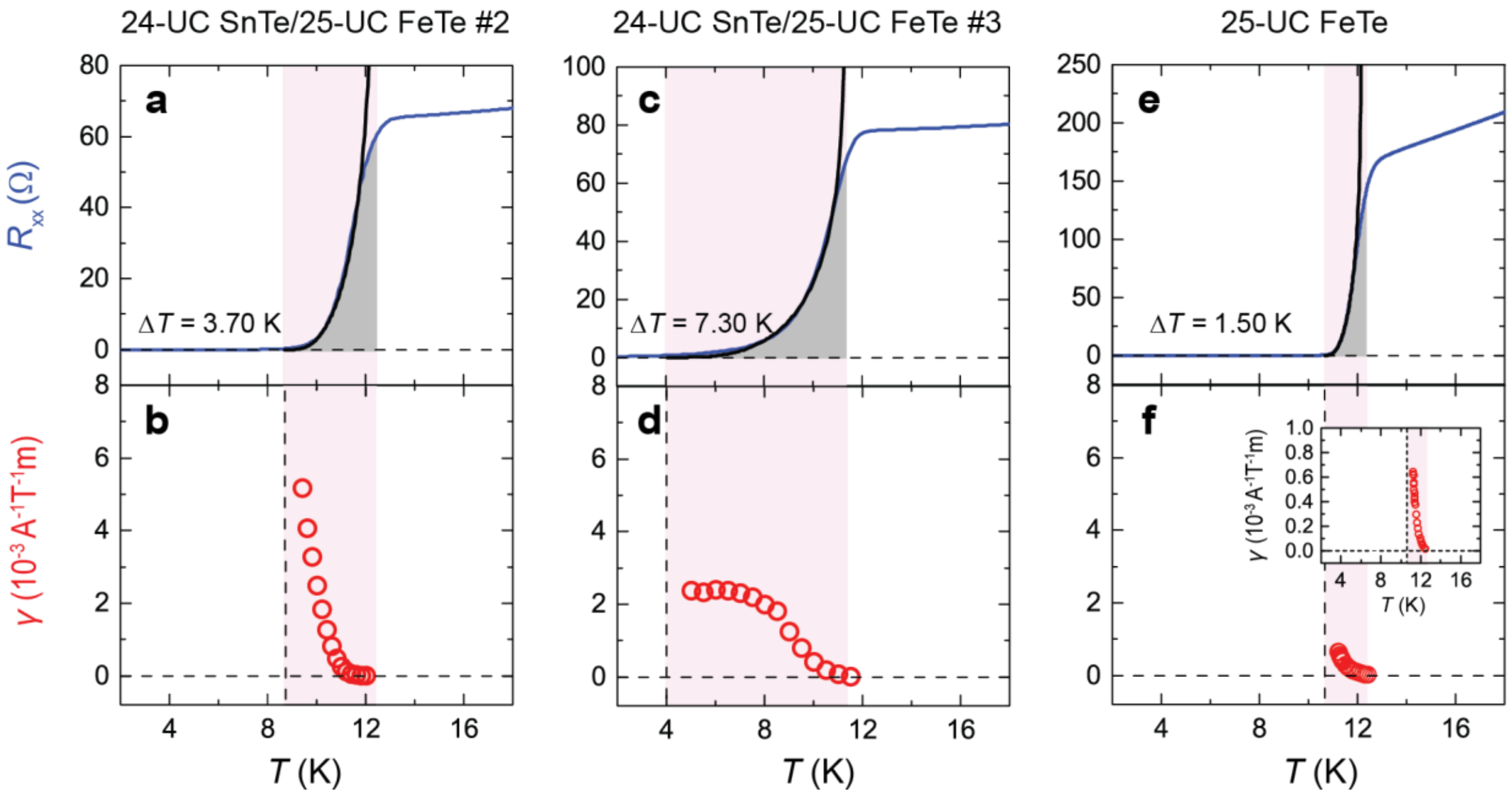

**Fig. 2 | Comparison of nonreciprocal charge transport in two additional 24-UC SnTe/25-UC FeTe heterostructures (Samples #2 and #3) and a 25-UC FeTe film**. **a**, **b**, $T$-dependent $R_{xx}$ (**a**) and $\gamma$ (**b**) in Sample #2 with $\Delta T = 3.70$ K. **c**, **d**, Same as (**a**, **b**), but for Sample #3 with $\Delta T = 7.30$ K. **e, f**, Same as (**a**, **b**), but for a 25-UC FeTe film with $\Delta T = 1.50$ K. Inset: Enlarged view of the $\gamma - T$ curve. The black curves in (**a, c, e**) are Halperin-Nelson fits used to determine $T_{\mathrm{BKT}}$. The pink shaded regions in (**a**-**f**) and the gray shaded regions under the $R_{xx}$-$T$ curve in (**a, c, e**) indicate the temperature range of $T_{\mathrm{BKT}} \leq T \leq T_{\mathrm{c,m}}$.

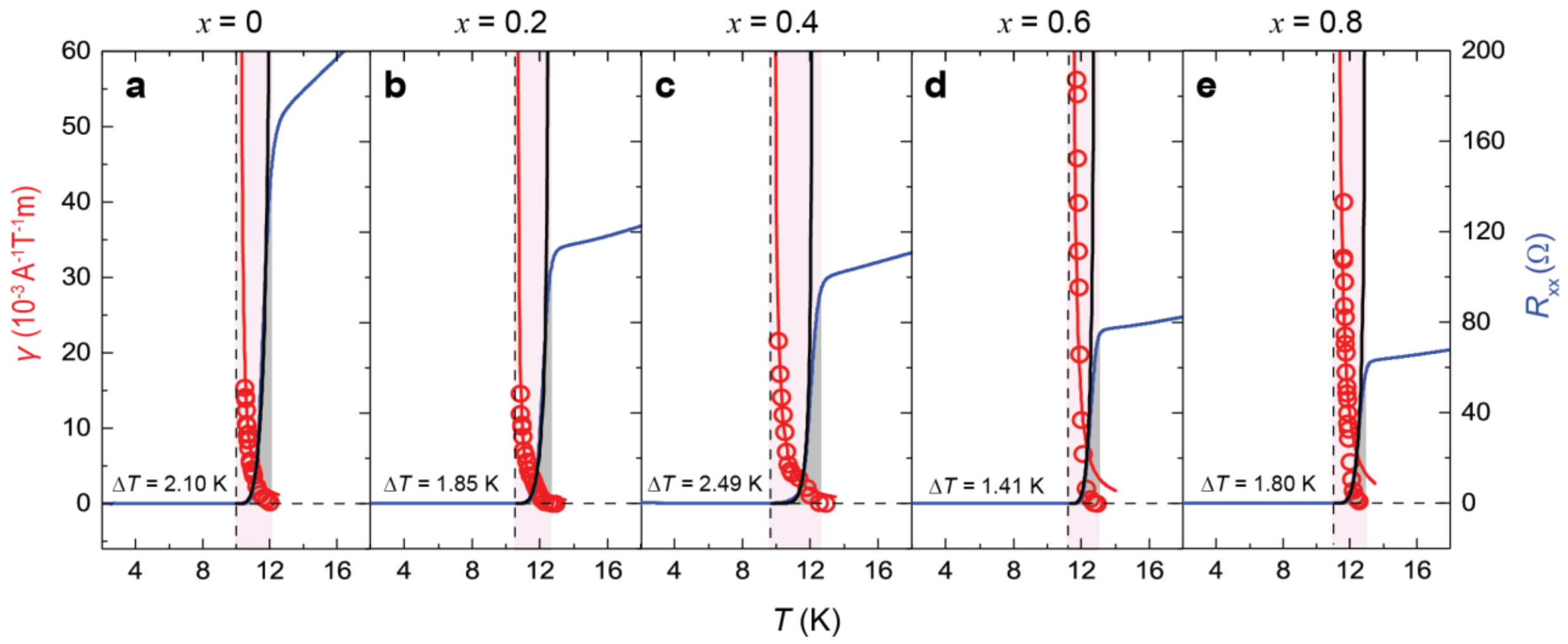

**Fig. 3 | Comparison of nonreciprocal charge transport in 24-UC $Pb_{1-x}Sn_xTe$/25-UC FeTe heterostructures with varying $x$.** **a**-**e**, $T$-dependent $\gamma$ (red circles) and $R_{xx}$ (blue curves) for the heterostructures with $x = 0$ (**a**), $x = 0.2$ (**b**), $x = 0.4$ (**c**), $x = 0.6$ (**d**), and $x = 0.8$ (**e**). The red curves in the $\gamma$-$T$ plots are fitted by the formula $\gamma = \gamma_0 (T - T_{BKT})^{-1.5}$ . The pink shaded regions in (**a**-**e**) and the gray shaded regions under the $R_{xx}$–$T$ curve in (**a**-**e**) indicate the temperature range of $T_{BKT} \leq T \leq T_{c,m}$.

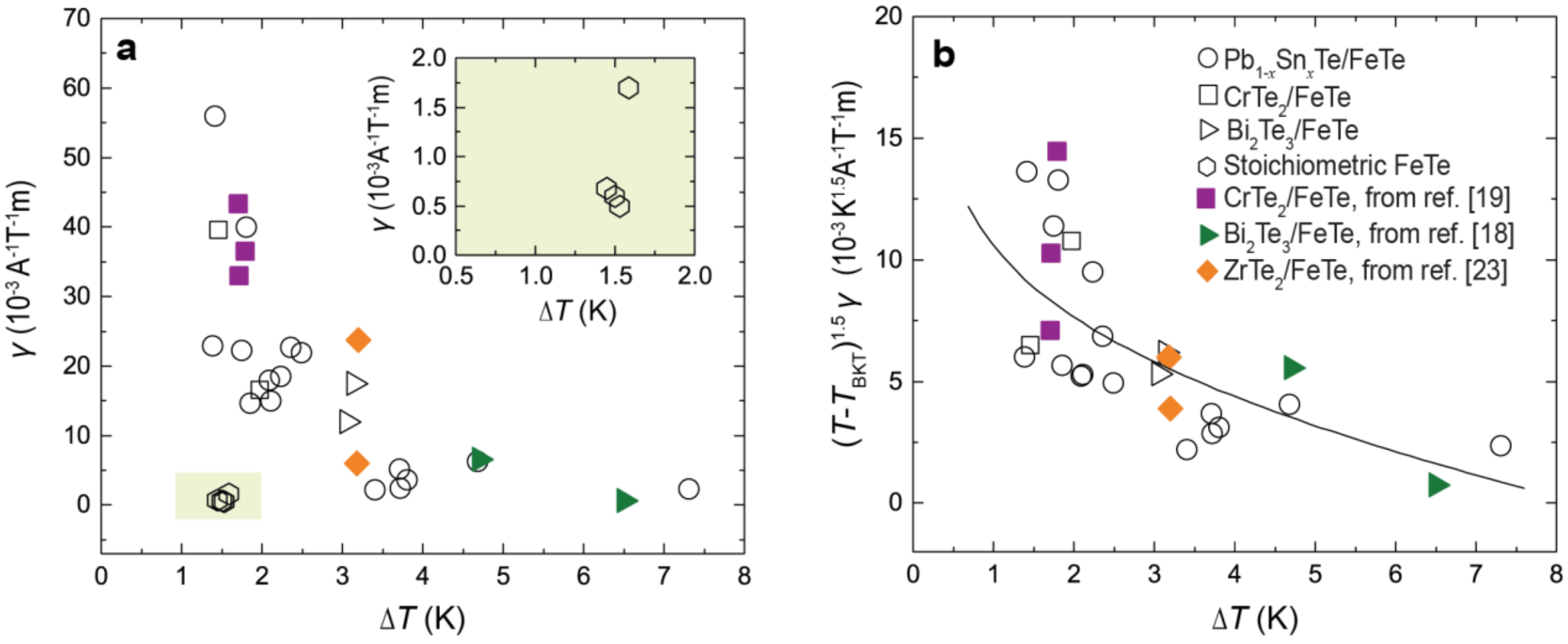

**Fig. 4 | Universal negative correlation between nonreciprocity and Δ*T* in superconducting FeTe-based heterostructures. a**, $\gamma$ as a function of $\Delta T$ for a broad range of superconducting FeTe-based heterostructures with different top-layer materials and layer thicknesses. For each sample, the reported $\gamma$ is the maximum value obtained from a linear fit to $R^{2\omega}/R^{\omega}$ versus $\mu_0 H_{||}$ with a coefficient of determination of $R^2 \geq 0.99$ at $T > T_{BKT}$. Inset: Enlarged view of the green-shaded region showing $\gamma$ as a function of $\Delta T$ for four control superconducting FeTe films with varying thicknesses. **b**, $(T - T_{BKT})^{1.5}\gamma$ as a function of $\Delta T$ for the same superconducting FeTe-based heterostructures shown in (**a**). The black curve represents the theoretically calculated dependence of $(T - T_{BKT})^{1.5}\gamma$ on $\Delta T$.